\documentclass[twocolumn,prl,showpacs,preprintnumbers,amsmath,amssymb]{revtex4}
\usepackage{graphicx}
\usepackage{dcolumn}
\usepackage{bm}
\usepackage{psfrag}
\usepackage{amsfonts}

\newcommand{\br}{\mathbf{r}}

\begin{document}

\preprint{APS/123-QED}

\title{Low complexity method for large-scale self-consistent \textit{ab initio} electronic-structure calculations
         without localization}

\author{M. J. Rayson}
 \affiliation{Institut f\"ur Physik, Universit\"at Basel, Klingelbergstrasse 82, 4056 Basel, Switzerland.}
\email{mark.rayson@unibas.ch}

\date{\today}

\begin{abstract}
A novel low complexity method to perform self-consistent
electronic-structure calculations using the Kohn-Sham
formalism of density functional theory is presented.
Localization constraints are neither imposed nor required
thereby allowing direct comparison with
\textit{conventional} cubically scaling
algorithms.  The method has, to date, the lowest complexity of any
algorithm for an \textit{exact} calculation.
A simple one-dimensional model system is used to thoroughly test
the numerical stability of the algorithm and results for
a real physical system are also given.
\end{abstract}


\maketitle
Low complexity electronic-structure methods \footnote{Throughout this letter 
it will be 
assumed the methods can furnish both the band-structure
energy \textit{and} the density leading to 
a self-consistent solution.},
using the Kohn-Sham density
functional approach \cite{bib:HK,bib:KS},   where the operation count
scales with respect to system size (`$N$-scaling') as
$N^\alpha$ 
 where $\alpha \leq 2$ 
have been around for a few decades.
A comprehensive review of low complexity
methods is given in reference \cite{bib:stefan-rmp}.
Contrary to what is often reported the theoretical
upper bound for the
$N$-scaling of an \textit{exact} self-consistent
algorithm has been set at $\mathcal{O}(N^2)$ ever since
Fermi operator expansion (FOE) methods were developed \cite{bib:stefan-poly1}.
This letter shows
how the theoretical upper bound for the scaling of such calculations
can be lowered to $\mathcal{O}(\alpha(d,N){N^\frac{2d-1}{d}} )$ where
$d$ is the \textit{dimensionality of repetition} of
a full three-dimensional system and $\alpha(1,N)=\log_2(N)$,
$\alpha(2,N) \leq 2$ and $\alpha(3,N) \leq 4/3$.  
For large scale calculations low complexity algorithms
are without doubt the future of electronic-structure
implementations. 
However, low complexity \textit{ab initio} algorithms
are not in common usage at the moment, primarily due 
to two main reasons.  Firstly, much of the work currently
being carried out deals with systems that are too small to
be amenable to low complexity approaches if high
accuracy is desired.  Secondly,
low complexity algorithms are not yet fully functional and fully
stable for general systems - so a sufficient level of confidence in using
these codes has not been established.  While the first
reason is rapidly being diminished due to the ever reducing cost
of a floating point operation, the second may prove to be far more 
stubborn.

Most low complexity algorithms fall broadly into two categories; 
either they attempt to calculate localized orbitals or they seek to 
evaluate the density matrix (DM) directly.  For a general
system only the latter is known to provide a low complexity solution.
In the case of a metal, for example, delocalized states at the Fermi
level prevent the occupied subspace being represented
in terms of orthogonal localized orbitals. 

Problems associated with low complexity approaches commonly
stem from the imposition of \textit{a priori} localization 
constraints.
The effect of this restriction varies depending on the
algorithm and physical system.  In orbital minimization algorithms
even the initial guess can alter the obtained solution.  In some cases localization will
always cast a degree of doubt over the final answers (except
in the simplest wide-gap systems), and
in others prohibits obtaining the relevant physics/chemistry
all together.
Fermi operator
expansion (FOE) algorithms (either using a polynomial \cite{bib:stefan-poly1,bib:stefan-poly2}
or rational \cite{bib:stefan-rational} approximation) for systems with a 
DM localized in real-space provide arguably the most natural and foolproof
way of obtaining results in $\mathcal{O}(N)$.  In these
methods the locality does not necessarily have to be imposed \textit{a priori}, rather the system can be allowed to inform us
of the locality in a systematic way.
Methods that impose unsystematic localization 
are invariably open to more doubt.  While a great deal of progress
has been made in understanding the inherent locality present in
many systems, low temperature metallic systems and charged insulating
systems with 
long-ranged DM correlations are still a significant challenge.
The method presented in this letter is primarily aimed
at such systems.
However, it has also been noted that
the onset of sparsity of the DM, even for wide-gap systems,
is `discouragingly slow' \cite{bib:Maslen} especially if high accuracy is required.
The main advantage of the method in this work is that it
relies purely on the locality of the basis functions allowing the use
of non-orthogonal localized basis sets, such as Gaussians, 
with rather less localized
orthogonal and dual complements. 
Also,
the full DM need not be explicitly calculated.

The energy renormalization group (ERG) approach
\cite{bib:ERG-Baer-1,bib:ERG-Baer-2,bib:ERG-kenoufi} is a beautiful and elegant
concept that has also been suggested to cope
with such difficult problems.  In an ideal implementation it may be 
possible for its scaling to better the method given here for $d>1$
and equal it for $d=1$.  However, it remains unclear
whether an ERG algorithm can also provide
the density in an efficient manner and to some extent the ERG
method employs cutoffs.  Therefore, the ERG method will
not be included in the definition of FOE methods in the 
following.

To date, standard FOE methods have been considered to scale quadratically
for systems where the DM decay length
is of the order of the system size.  This can be the case for 
very large systems especially for metals at low
temperature or if high accuracy is required.  The method presented here imposes no localization
constraints and scales as $\mathcal{O}(\alpha(d,N){N^\frac{2d-1}{d}} )$
where $\alpha(d,N)$ is a weak logarithmic factor for $d=1$ and
tends to a constant in higher dimensions.
Not only does this represent
a new theoretical upper bound for the $N$-scaling of an
\textit{exact} algorithm (upto the basis set limit) it is also
expected to make a significant and
immediate impact on 
systems of low dimensionality.
Furthermore, for $d=1$ it can be implemented using 
exclusively standard direct linear algebra routines (eg. LAPACK)
for the bulk of the computation.  This is because a $d=1$ 
Hamiltonian (with zero or periodic boundary conditions)
can always be arranged so that it is a banded matrix, with
a bandwidth that is independent of system size, if it is
constructed from localized basis functions.

We now turn to what will be referred
to as the recursive bisection density matrix
(RBDM) algorithm.
We begin with a rational approximation of the density matrix
 \footnote{A polynomial expansion
could also be used.} 
\begin{equation}
  F(H) = \frac{1}{1+e^{\beta(H-\mu)}}\simeq \sum_k^{n_\mathrm{r}} \omega_k (H-z_k)^{-1} ~~~ \omega_k, z_k \in \mathbb{C}, \label{eqn:ratapp}
\end{equation}
where $\beta$ and  $\mu$ are the inverse temperature and
Fermi energy respectively.  
The inverses of the shifted Hamiltonians in equation~(\ref{eqn:ratapp}) may be 
evaluated by solving linear equations.  A number of methods
to construct such rational approximations have previously
appeared in the literature \cite{bib:ratapp-stefan,bib:ratapp-nicholson,bib:ratapp-gagel}.
For a given temperature,
the condition of the shifted matrices is asymptotically independent of system size.
Therefore, if no localization of the DM can be taken
advantage of the solution of each equation requires $\mathcal{O}(N)$ operations.
Since we must solve $\mathcal{O}(N)$ equations
the overall scaling is $\mathcal{O}(N^2)$ - as stated previously.
A key point is that
to calculate the band-structure energy and density
\begin{equation}
  E_{\mathrm{bs}} = \sum_{ij} F_{ij}H_{ij}, ~~ n(\br) = \sum_{ij} F_{ij} \phi_i(\br)\phi_j(\br)
\end{equation}
using a localized basis set $\{\phi_i\}$ only requires elements of the DM that lie within the sparsity
pattern of the Hamiltonian.
The inverse of such a shifted matrix is
clearly symmetric as
\begin{eqnarray}
  (H-z_k)^{-1} = c \Lambda c^T, ~~~ \Lambda_{ij} = \delta_{ij}/(\lambda_i-z_k) \label{eqn:symm1} \\
  {[(H-z_k)^{-1}]}^T  = {[c \Lambda c^T]}^T = [(\Lambda c^T)^T c^T] = c \Lambda c^T
\label{eqn:symm2}
\end{eqnarray}
where $\{\lambda_i\}$ and $c$ are the eigenvalues and eigenvectors of $H$
respectively.
For simplicity the Hamiltonian matrix
$H$ in equations~(\ref{eqn:ratapp})-(\ref{eqn:symm2}) is taken to be constructed from an orthogonal
basis.  Generalization to the non-orthogonal case simply
requires replacing $(H-z_k)$ with $(H-z_k S)$ throughout
and noting that the eigenvectors in equations (\ref{eqn:symm1})
and (\ref{eqn:symm2}) satisfy $c^TSc = I$ where $S$ is the 
overlap matrix of the basis functions.

We may then proceed with a recursive bisection of the matrix approach
without approximation.
The easiest way to demonstrate this principle is to
see how one can obtain the density for a 
$d=1$ system, such as a linear molecule
or carbon nanotube.
For such a system the Hamiltonian is a banded
matrix.  The width of the band, although independent of system size, is 
implementation and system specific.  Therefore,
for the sake of clarity a truly one-dimensional
system will be considered.
The simplest Hamiltonian we can imagine is a
finite-difference stencil representing
the Laplacian and the local potential represented
on a grid of spacing $h$
\begin{eqnarray}
  H_{ii} &=& 1/h^2 + V(x_i) \nonumber \\
  H_{ij} &=& -1/(2h^2),  |i-j|=1 \label{eqn:simpham} \\
  H_{ij} &=& 0,   \nonumber |i-j|>1.
\end{eqnarray}
As this matrix is tridiagonal, a submatrix (on the diagonal) of $H$
requires two boundary points to determine the linear
equation $(H_\mathrm{sub} - z_k)x=b$.
\begin{figure}
\begin{center}
  \includegraphics[width=3.375in]{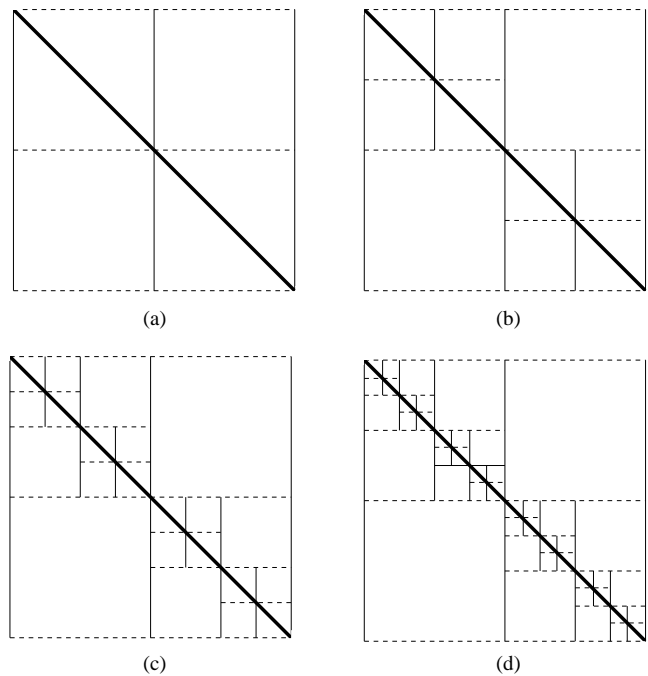}
 \caption{\label{fig:dcdm} Schematic of bisection of matrix inverses for $d=1$.
       The broad diagonal line represents the band of the matrix.
   The narrow vertical lines represent the columns
   of the matrix inverses $(H-z_k)^{-1}$
 that are calculated.  The dashed horizontal lines are rows that are
known from the calculated columns due to the matrix being symmetric.  These
rows then specify boundary conditions for smaller sets of independent linear
equations at each \textit{sweep} (a-d).}
\end{center}
\end{figure}
Fig. \ref{fig:dcdm} shows a schematic of the RBDM strategy 
for $d=1$.  After the first \textit{sweep} (Fig 1 (a)) the first,
\textit{central} and last columns are known.
From the rows (known as the matrix is symmetric) we now have boundary conditions of two smaller
problems which can be solved independently (Fig 1 (b)).  We may then bisect these two subproblems 
in a similar fashion (Fig 1 (c)).  The process continues until the dimensions
of the submatrices are comparable to the bandwidth
of the matrix (Fig 1(d)), and then direct evaluation can be used for
the remaining subproblems (smallest blocks on the diagonal in
 Fig 1(d)).

We now turn to the scaling of the method for $1 \geq d \leq 3$.
We start with a cubic system and imagine increasing
the size of the system by a factor $\gamma$ in
$d$ dimensions thereby increasing the total size
of the system by $\gamma^d$.  Firstly, we consider
only the cost of the first bisection (Fig 1 (a)) of the system and
we consider the DM to have effectively infinite range.
To bisect the system into two subsystems requires
calculating $n_\mathrm{col}$ columns (represented by vertical lines
in Fig 1) of the DM and each column
requires $\mathcal{O}(N)$ operations to compute.
As the system size is increased $\gamma^{(d-1)}n_\mathrm{col}$
columns are required to bisect the system.
Therefore, the first sweep scales as $\mathcal{O}(N^{(2d-1)/d})$ - 
and this is the leading term.  
This bisection operation must then be repeated until
all of the desired elements of the DM have been calculated.
The number of bisections required
goes like $\log_{2^d} (N)$.  The number of operations
required to perform sweep $m$ ($m>1)$ is $\sim N_1/2^{(d-1)(m-1)}$ where
$N_1$ is the number of operations to perform the first sweep.
Therefore, the total number of operations may be
written as
\begin{equation}
  N_\mathrm{tot} \propto N^{(2d-1)/d} \sum_{m=0}^{\sim\log_{2^d}(N)} \frac{1}{2^{(d-1)(m-1)}}.
\end{equation}
For $d=1$ the summation is clearly proportional to $\log_2 N$.  However,
in higher dimensions the summation is a convergent series and gives
2 for $d=2$ and $4/3$ for $d=3$.  This is an upper bound for the
number of operations.  Elaborate bisection schemes may reduce
the total number of operations but the leading scaling with
$N$ will not be affected.
Hamiltonians with broader bands from the use of more
extended basis functions or non-local pseudopotentials
require an increase in $n_\mathrm{col}$, however,  this
does not affect the $N$-scaling.

Another important aspect of any algorithm is numerical
stability. 
As many elements of the DM rely on previous 
solutions of linear equations we may expect errors
to accumulate the more bisections we use.  It is
difficult to gauge the precise effect on the total energy,
however we may concentrate on a single inverse and assume
the worst case scenario.  If we take one of
our shifted matrices that is closest to being
singular (the matrix shifted closest to the Fermi energy)
$(H-z_\mathrm{c})$ then the error in solving for 
one column of the matrix is proportional to $\epsilon_\mathrm{m}\kappa(H-z_\mathrm{c})$
where $\epsilon_\mathrm{m}$ and $\kappa$ are machine precision and condition
of the matrix respectively.  At worse we may expect the error to 
grow linearly with the bisection number, though a random-walk
accumulation leading to a square root dependence
is more realistic.  Fig. 2 shows this
slow \textit{drift} in the value of $Tr(H(H-z_\mathrm{c})^{-1})$
where $(H-z_\mathrm{c})^{-1}$ is a very ill conditioned
matrix (certainly as ill-conditioned as any in a realistic
electronic structure calculation).  However, each
submatrix will have eigenvalue range
similar to that of the full matrix but a less
clustered eigenspectrum.  This will render
\textit{sub}-linear systems becoming further from
singularity during the bisection process.  The numerics in a full calculation
are clearly very complex.
One-dimensional
model systems were extensively tested in single precision,
 including double precision iterative improvement of
the solutions, from a range
of ill-conditioned matrices. In some cases
increasing the bisection number produced
results closer to that of double precision diagonalization
and no catastrophic numerical instabilities
were detected.
\begin{figure}
\begin{center}
  \includegraphics[width=3.375in]{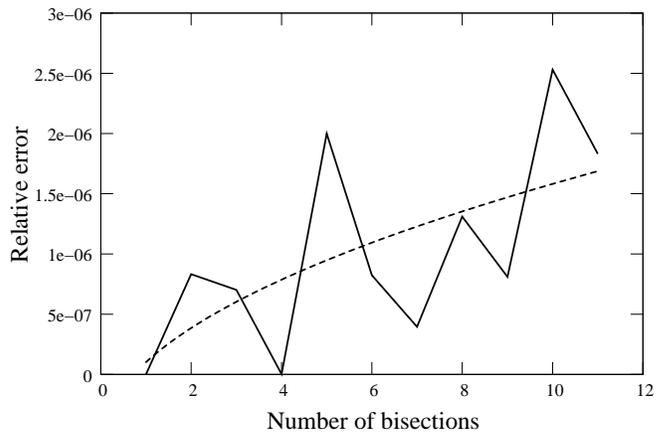}
 \caption{Relative error in $Tr(H(H-z_\mathrm{c})^{-1})$ (single precision) compared
   to the case where no bisections were used (solid line).  The condition of
  $(H-z_\mathrm{c})^{-1}$ is of the order $10^6$.
 For such an ill-conditioned matrix even the relative error
 in single precision direct diagonalization was $\sim 10^{-5}$.
The dashed line shows a fit of
 the square root of the number of bisections.}
\end{center}
\end{figure}

As a final example we take a more physically
realistic Hamiltonian.  A minimal Gaussian
basis was used to construct Hamiltonian and
overlap matrices for linear C$_\mathrm{n}$H$_{\mathrm{2n+2}}$
molecules using a norm-conserving non-local
pseudopotential \cite{bib:HGH-PP}.  To obtain a physically
reasonable eigenspectrum using the minimal basis
for this molecule requires
basis functions with a spatial extent which corresponds to
the bandwidth of the matrix being approximately 50.  This 
corresponds to a chain length of around 8 carbon
atoms before the bandwidth of the matrix becomes
less than the dimension of the matrix.  
For testing purposes a 
low temperature ($\sim 0.04$eV ) Fermi distribution
distribution with $\mu$ taken
to be an eigenvalue in the valence
band was chosen.  This corresponds to
a highly charged insulating system with
a long range DM (Fig. 3) and also provides
an ill-conditioned problem ideal to test numerical
stability. 
The absolute/relative error, compared to direct diagonalization,
for the 1001 atom C$_{333}$H$_{668}$ was
$\sim 10^{-10}$/$10^{-13}$ and 5 bisections were required.
This further puts into context the numerical drift
mentioned in the previous section.  No iterative
improvement was used in this example, only full 
double precision arithmetic, and the 
ill-conditioning of the linear systems represents
the worst case in a typical calculation.
Therefore, in a realistic calculation,
chain lengths containing at least one million
basis functions in one-dimension (and more in higher
dimensions) should be accessible 
(by which point the natural decay of the density matrix
will surely limit the number of required bisections in any case).
\begin{figure}
\begin{center}
  \includegraphics[width=3.375in]{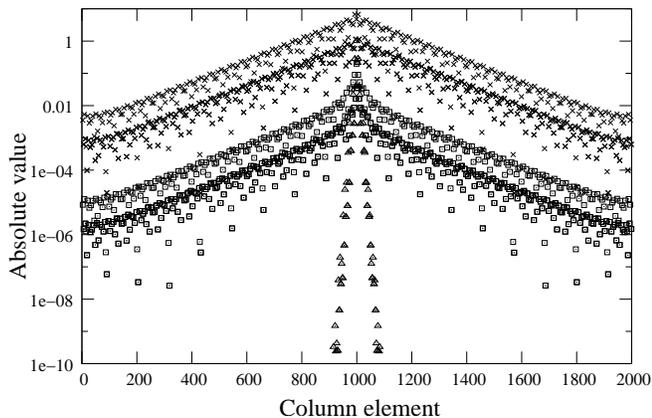}
 \caption{Decay of the central column of the density
 matrix (squares) and two inverses of shifted Hamiltonians
for the
 highly charged C$_{333}$H$_{668}$ system.
   The Hamiltonian
 closest to the Fermi energy (crosses) and the Hamiltonian
shifted furthest from the real axis (triangles).}
\end{center}
\end{figure}

We now discuss some further implementation issues.
For large systematic basis sets the memory required to store
the boundary conditions may become 
prohibitive - especially in three dimensions.  The method
can overcome this to some extent by bisecting
the system by a factor, $q$, greater than two and building
up the density matrix in segments. 
However, when using large 
basis sets, a smaller \textit{filtered} set of basis functions
expanded in terms of the underlying basis would be a more
realistic approach.
It can now be clearly seen how conventional linear algebra 
can be used for $d=1$ systems.  A banded matrix
can be $LU$ factorized in $\mathcal{O}(N)$ operations
and a linear equation solved in $\mathcal{O}(N)$ using direct
methods.  
Therefore, for $d=1$ iterative algorithms
need not be considered - this is useful when using
localized basis functions such as Gaussians where iterative methods
are still difficult to precondition. 
Also, the matrices shifted close to $\mu$, at low
temperature, become close to singular therefore even basis sets
that can be readily preconditioned in a \textit{conventional}
 sense (by damping of high kinetic energy components) will
also suffer in this regime, so direct methods are desirable.
As solving sparse linear systems of equations forms the kernel of the method it
is naturally open to any advances in direct sparse solvers for
systems where $d>1$.

In principle, a similar procedure can be used 
if one opts for a polynomial, rather than a rational, approximation
to the Fermi function.  
If $F(H)$ is approximated by a polynomial in $H$, 
$F(H) \simeq   \sum_k^{n_\mathrm{p}} \omega_k H^k$, we may construct
a set of columns of $H^k [k=2,...,n_\mathrm{p}]$ and store
the necessary \textit{boundary matrix elements} for each
$k$ in a similar fashion to that already described above.

Even if a system has a DM that is sufficiently localized
to take advantage of the RBDM method can still
be used to dramatically reduce the
prefactor if the localization regions 
are significantly larger than the spatial extent
of the basis functions.  This will often be the case
if highly accurate relative energies are desired.  Also,
the inverses of Hamiltonians shifted far from the real-axis
have more rapid decay allowing true $\mathcal{O}(N)$ evaluation
(Fig. 3).

In conclusion, a simple modification of FOE methods
has been presented allowing 
$\mathcal{O}(\alpha(d,N){N^\frac{2d-1}{d}} )$ scaling where
$\alpha(1,N)=\log_2(N)$,
$\alpha(2,N) \leq 2$ and $\alpha(3,N) \leq 4/3$ without
the need for localization.
This is a especially useful for systems of low dimensionality
with long-ranged DM correlations.

The author thanks S. Goedecker
for helpful comments regarding the manuscript and
P. R. Briddon for providing the C$_{333}$H$_{668}$ test matrix.
This work was
supported by the European Commission within the Sixth
Framework Programme through NEST-BigDFT 
(Contract
No. BigDFT-511815).

\bibliography{text.bib}

\end{document}